\documentclass{appolb}
\usepackage{graphicx}

\begin{document}
\title{Search for $\eta'$(958)-nucleus bound states by $(p,d)$ reaction at GSI and FAIR%
\thanks{Presented at II Symposium on applied nuclear physics and innovative technologies.}%
}
\author{H.~Fujioka$^{a}$, Y.~Ayyad$^{b}$, J.~Benlliure$^{c}$, K.-T.~Brinkmann$^{d}$,
S.~Friedrich$^{d}$, H.~Geissel$^{d,e}$, J.~Gellanki$^{f}$, C.~Guo$^{g}$, E.~Gutz$^{d}$,
E.~Haettner$^{e}$, M.N.~Harakeh$^{f}$, R.S. Hayano$^{h}$, Y.~Higashi$^{i}$, S.~Hirenzaki$^{i}$, C.~Hornung$^{d}$, Y.~Igarashi$^{j}$, N.~Ikeno$^{k,l}$, K.~Itahashi$^{m}$, M.~Iwasaki$^{m}$, D.~Jido$^{n}$, N.~Kalantar-Nayestanaki$^{f}$, R.~Kanungo$^{o}$, R.~Knoebel$^{d,e}$,
N.~Kurz$^{e}$, V.~Metag$^{d}$, I.~Mukha$^{e}$, T.~Nagae$^{a}$, H.~Nagahiro$^{i}$,
M.~Nanova$^{d}$, T.~Nishi$^{h}$, H.J.~Ong$^{b}$, S.~Pietri$^{e}$, A.~Prochazka$^{e}$, 
C.~Rappold$^{e}$, M.P. Reiter$^{e}$, J.L.~Rodr\'{i}guez-S\'{a}nchez$^{c}$, C.~Scheidenberger$^{d,e}$,
H.~Simon$^{e}$, B.~Sitar$^{p}$,P.~Strmen$^{p}$, B.~Sun$^{g}$, K.~Suzuki$^{q}$, 
I.~Szarka$^{p}$, M.~Takechi$^{r}$, Y.K. Tanaka$^{h}$, I.~Tanihata$^{b,g}$, S.~Terashima$^{g}$,
Y.N.~Watanabe$^{h}$, H.~Weick$^{e}$, E.~Widmann$^{q}$, J.S.~Winfield$^{e}$, X.~Xu$^{e}$, H.~Yamakami$^{a}$, J.~Zhao$^{g}$\\for the Super-FRS Collaboration
\address{
$^{a}$Kyoto University, Kitashirakawa-Oiwakecho, Sakyo-ku, 606-8502 Kyoto, Japan\\
$^{b}$RCNP, Osaka University, 10-1 Mihogaoka, Ibaraki, 567-0047 Osaka, Japan\\
$^{c}$Universidade de Santiago de Compostela, 15782 Santiago de Compostela, Spain\\
$^{d}$Universit\"{a}t Giessen, Heinrich-Buff-Ring 16, 35392 Giessen, Germany\\
$^{e}$GSI, Planckstrasse 1, 64291 Darmstadt, Germany\\
$^{f}$KVI-CART, University of Groningen, Zernikelaan 25, 9747 AA Groningen, the Netherlands\\
$^{g}$Beihang University, Xueyuan Road 37, Haidian District, 100191 Beijing, China\\
$^{h}$The University of Tokyo, 7-3-1 Hongo, Bunkyo, 113-0033 Tokyo, Japan\\
$^{i}$Nara Women's University, Kita-Uoya Nishi-Machi, 630-8506 Nara, Japan\\
$^{j}$KEK, 1-1 Oho, Tsukuba, 305-0801 Ibaraki, Japan\\
$^{k}$Tohoku University, 6-3 Aoba, Aramaki, Aoba, Sendai, 980-8578 Miyagi, Japan\\
$^{l}$YITP, Kyoto University, Kitashirakawa-Oiwakecho, Sakyo-ku, 606-8502 Kyoto, Japan\\
$^{m}$Nishina Center, RIKEN, 2-1 Hirosawa, Wako, 351-0198 Saitama, Japan\\
$^{n}$Tokyo Metropolitan University, 1-1 Minami-Osawa, Hachioji, 192-0397 Tokyo, Japan\\
$^{o}$Saint Mary's University, 923 Robie Street, Halifax, Nova Scotia B3H 3C3, Canada\\
$^{p}$Comenius University Bratislava, Mlynsk dolina, 842 48 Bratislava, Slovakia\\
$^{q}$Stefan-Meyer-Institut f\"{u}r subatomare Physik, Boltzmangasse 3, 1090 Vienna, Austria\\
$^{r}$Niigata University, 8050 Ikarashi 2-no-cho, Nishi-ku, 950-2181 Niigata, Japan}
}

\maketitle
\begin{abstract}
The mass of the $\eta'$ meson is theoretically expected to be reduced at finite density,
which indicates the existence of $\eta'$-nucleus bound states.
To investigate these states, we perform missing-mass spectroscopy for the $(p,d)$ reaction 
near the $\eta'$ production threshold.
The overview of the experimental situation is given and the current status is discussed.
\end{abstract}
\PACS{21.85.+d, 25.40.Ve}

%
%
\section{Introduction}
The $\eta'$ meson, one of the pseudoscalar mesons, is known to have a peculiarly large mass ($958\,\mathrm{MeV}/c^2$).
It is attributed to the $U_A(1)$ anomaly in Quantum Chromodynamics, together with
the spontaneous breaking of chiral symmetry~\cite{Jido12}.
At finite density, in which chiral symmetry will be partially restored, 
the mass is expected to be decreased.
It means that an $\eta'$ meson in a nucleus feels an attractive interaction with the nucleus,
and that they may form a bound state ($\eta'$-mesic nucleus),
if the mass reduction is large enough. For example,
calculations based on the Nambu--Jona-Lasinio model result in a very large mass reduction of $-150\,\mathrm{MeV}$ at the normal nuclear density~\cite{Nagahiro06}, 
while a recent calculation using the linear sigma model shows $-80\,\mathrm{MeV}$ mass reduction~\cite{Sakai13}. The QMC (quark-meson coupling) model predicts $-37\,\mathrm{MeV}$ reduction
for the case of an $\eta$-$\eta'$ mixing angle of $-20^\circ$~\cite{Bass06}.

As for the decay width of an $\eta'$-mesic nucleus, the chiral unitary model including a Lagrangian term which couples the singlet meson to the baryons indicates that the real part of the optical potential is in general deeper than the imaginary part~\cite{Nagahiro12}. This relationship is one of the important requirements for an $\eta'$-mesic nucleus to be observed experimentally.

Recently, the CBELSA/TAPS experiment determined the real part~\cite{Nanova13} and the imaginary part~\cite{Nanova12} through detailed studies of the $\eta'$ photoproduction off nuclei.
First, the transparency ratios for different nuclear targets and incident energies were compared with
theoretical calculations, and the absorption width for an $\eta'$ meson with an average momentum of $1050\,\mathrm{MeV/}c$ was obtained to be $15\mbox{--}25\,\mathrm{MeV}$ at the normal nuclear density~\cite{Nanova12}.
Furthermore, the excitation function for $\eta'$ photoproduction off carbon and
the momentum distribution of $\eta'$ meson are sensitive to the $\eta'$-nucleus potential depth.
The potential depth was derived as $-(37\pm 10\mbox{(stat)} \pm 10\mbox{(syst)})\,\mathrm{MeV}$~\cite{Nanova13}. The combination of the two results indicates
the possible existence of $\eta'$-mesic nuclei with a rather narrow width.
Related to the study of $\eta'$-nucleus interaction, the $\eta'$-nucleon scattering length
can be determined by the measurement of the $pp\to pp\eta'$ reaction~\cite{Czerwinski14}.
The result is comparable to the $\eta'$-nucleus optical potential parameters obtained by CBELSA/TAPS,
if there is no density and energy dependence of the $\eta'$-nucleon interaction.

Another method to extract information on the $\eta'$-nucleus interaction is the investigation of $\eta'$-mesic nuclei. We have proposed a missing-mass spectroscopy experiment of $\eta'$-mesic nuclei by use of the $(p,d)$ reaction~\cite{Itahashi12}. The first experiment was carried out at GSI in August 2014, and an upgraded experiment at FAIR is also planned.

\section{Inclusive measurement at GSI}
$\eta'$-mesic nuclei can be produced by impinging a proton beam onto a carbon target.
The beam energy is chosen to be $2.5\,\mathrm{GeV}$, slightly above the $\eta'$ production threshold for a free nucleon. The ejectile deuterons will be momentum-analyzed by the Fragment Separator (FRS) used as a high-resolution spectrometer. 

\begin{figure}[t]
\begin{center}
\resizebox{0.75\columnwidth}{!}{%
\includegraphics{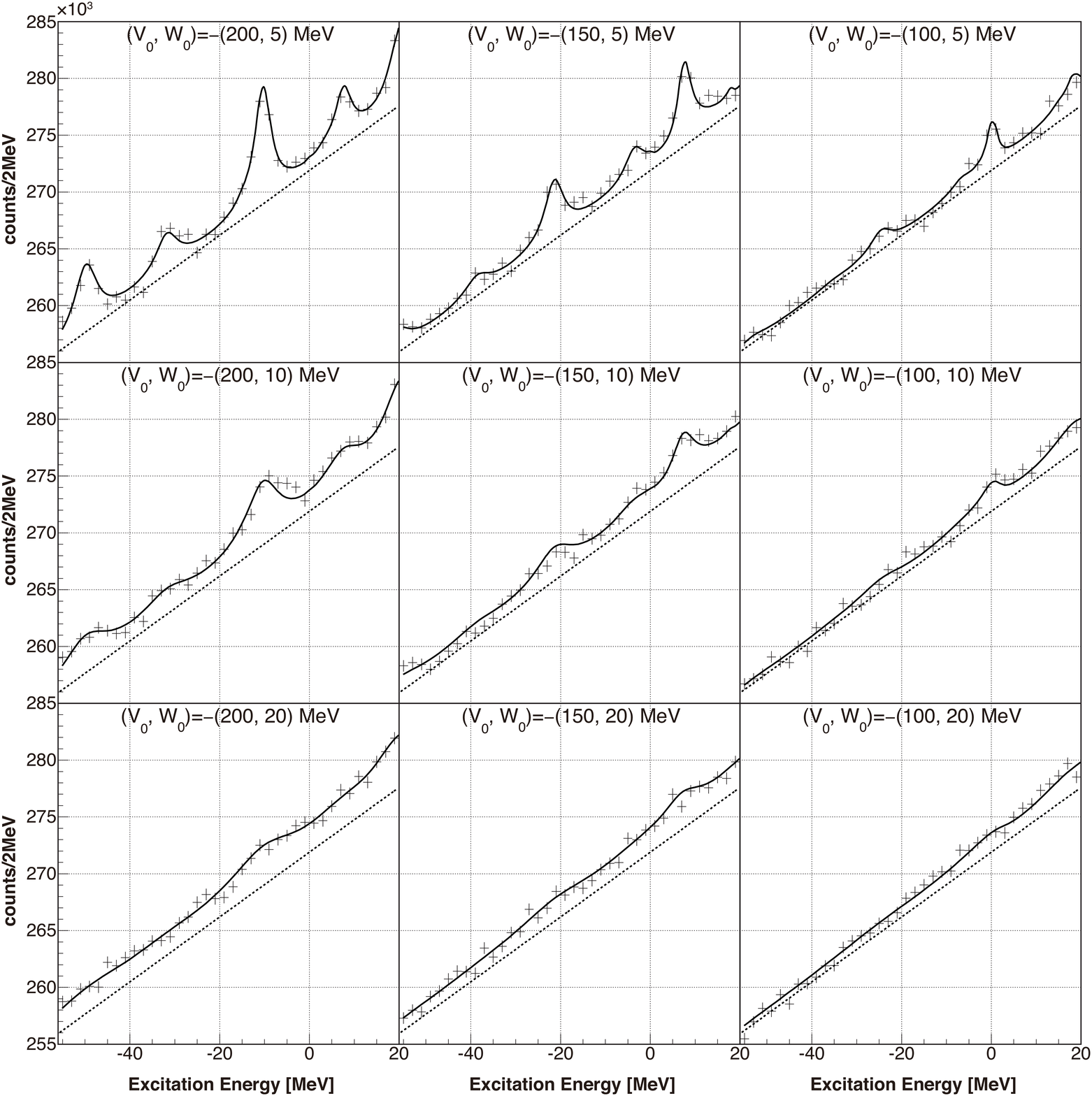} }
\caption{Simulated spectra with 4.5 days of data acquisition for different $\eta'$-nucleus optical potentials, parameterized as $V_0 + iW_0$.
The dashed line corresponds to the background of quasi-free multi-pion production processes.}
\label{missingmass}       
\end{center}
\end{figure}

A simulated spectrum corresponding to 4.5 days of data acquisition is shown in Fig.~\ref{missingmass}.
Here, a theoretical calculation for this reaction~\cite{Nagahiro13} is used as the input of the signal, i.e.~processes associated with $\eta'$ meson production, while an overwhelming background of multi-pion production processes on a nucleon in the carbon nucleus (mainly $p+N\to d+2\pi$, $3\pi$, or 4$\pi$) was evaluated from past measurements of proton-nucleon cross sections. The sensitivity of observing peak structures is highly dependent on the parameters of the optical potential ($V_0+iW_0$). We find the signal-to-noise ratio to be of the order of 1/100 at most. However, the poor signal-to-noise ratio in the inclusive measurement may be compensated by a high-statistics measurement for several days.
 
\begin{figure}[ht]
\begin{center}
\resizebox{0.7\columnwidth}{!}{%
\includegraphics{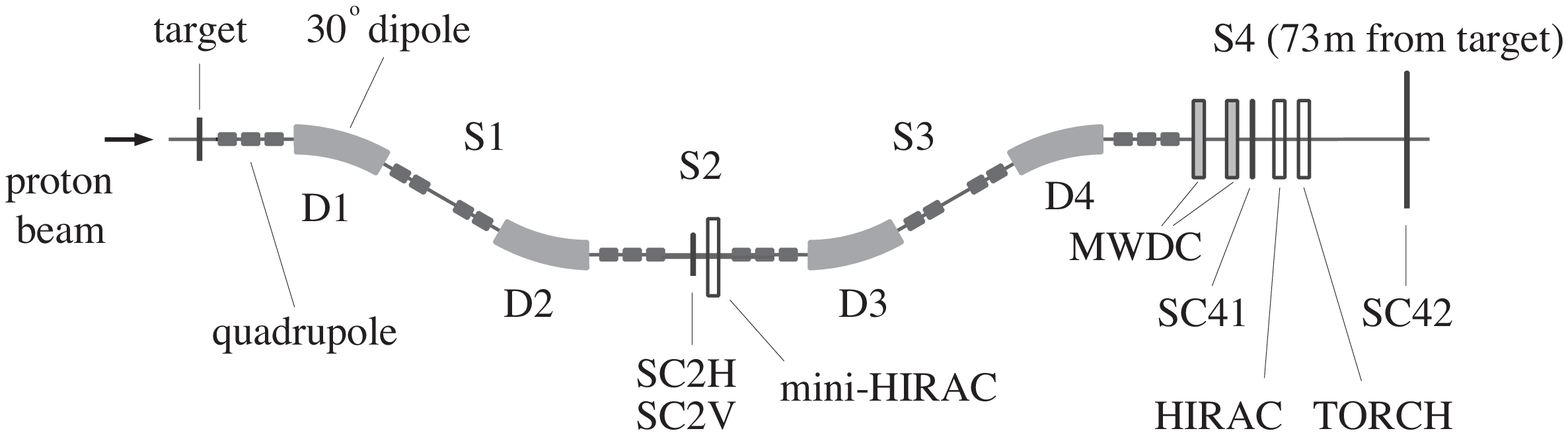} }
\caption{Schematic view of the experimental setup in August 2014.}
\label{frs_setup}       
\end{center}
\end{figure}

We carried out the first experiment (GSI S-437) in August 2014.
Figure~\ref{frs_setup} shows the experimental setup at FRS.
The momentum of ejectile deuterons can be derived by measuring their tracks by two sets of multi-wire drift chambers (MWDC's) installed near a dispersive focal plane at S4.
The overall resolution had been evaluated to be approximately $1.6\,\mathrm{MeV}$ in $\sigma$, which is
sufficiently smaller than the expected width of $\eta'$-mesic nuclei.
The actual resolution will be evaluated
by using the calibration measurement of the elastic scattering $D(p,d)p$ reaction.
The excitation energies between $-90\,\mathrm{MeV}$ and $40\,\mathrm{MeV}$ relative to the $\eta'$ emission threshold were investigated by scaling the magnetic field of FRS.

Particle identification of the ejectiles is essential,
since inelastic scattering $(p,p')$ reactions will cause a number of protons 
whose momenta are close to those of deuterons of interest.
For this purpose, plastic scintillators (SC2H, SC2V, SC41, SC42) were installed both in the S2 and S4 areas.
In addition to the time-of-flight (TOF) measurement, \v{C}erenkov detectors with high refractive-index ($n= 1.17\mbox{--}1.18$) silica aerogel radiators, which were developed at Chiba University~\cite{Tabata}, (HIRAC and mini-HIRAC) and a total-reflection \v{C}erenkov detector with an Acrylite radiator (TORCH) were installed 
for additional information on particle identification.

\begin{figure}[t]
\begin{center}
\resizebox{0.45\columnwidth}{!}{%
\includegraphics{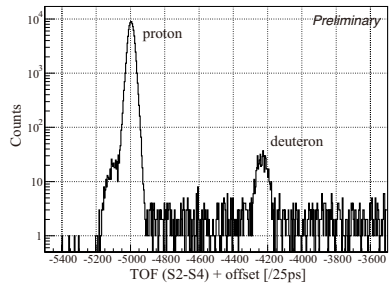} }
\caption{Particle identification using time of flight between S2 and S4.}
\label{tof2h41}       
\end{center}
\end{figure}

Figure~\ref{tof2h41} is the TOF distribution between S2 and S4 with an unbiased trigger.
The larger peak corresponds to protons, which are faster than deuterons, and the smaller peak comes from deuterons. The TOF difference of the two particles is about $20\,\mathrm{ns}$, which is consistent
with that calculated from the flight distance and each velocity. The deuteron-to-proton ratio is found to be approximately 1/200. After this measurement, we could prepare a ``TOF trigger'' to select deuterons by a tight coincidence of signals from scintillators at S2 and S4. We realized much better deuteron-to-proton ratio, approximately unity, without using the signals from the \v{C}erenkov detectors, while they can be used for improving off-line analyses. 

At present, we have been working on off-line analyses for particle identification. The flat component beneath the deuteron peak in Fig.~\ref{tof2h41} is due to sequential protons with a short ($\sim 20\,\mathrm{ns}$) time interval. The waveforms of the scintillator signals will serve to distinguish one-pulse events from two-pulse events.

\section{Semi-exclusive measurement at FAIR}
In order to improve the signal-to-noise ratio in the missing-mass spectrum,
we plan to measure protons from the decay of $\eta'$-mesic nuclei in addition to ejectile deuterons.
One of the major decay modes is a two-nucleon absorption process ($\eta'NN\to NN$)~\cite{Nagahiro13},
which will emit a proton with a kinetic energy around $300\mbox{--}600\,\mathrm{MeV}$.
An intra-nuclear cascade simulation with a microscopic transport model JAM~\cite{JAM} is on-going,
not only for the signal but also the background multi-pion production processes, in which pions may result in the emission of secondary protons after rescattering inside the nucleus.

\section{Summary}
We have performed an inclusive measurement of the $(p,d)$ reaction on ${}^{12}\mathrm{C}$ at FRS/GSI in August 2014, in order to investigate $\eta'$-mesic nuclei. A wide excitation-energy range between $-90\,\mathrm{MeV}$ and $40\,\mathrm{MeV}$ was investigated. The expected resolution will be $\sigma\sim 1.6\,\mathrm{MeV}/c^2$, which is much smaller than the width of $\eta'$-mesic nuclei. The analysis is in progress.

Furthermore, we aim to conduct a semi-exclusive measurement at FAIR, in which protons from the two-nucleon absorption of an $\eta'$ meson in nucleus will be detected too. The signal-to-noise ratio in the missing-mass spectrum will be improved drastically by the coincidence measurement. A detailed simulation taking into account the final-state interaction is needed for a quantitative evaluation of the sensitivity.
\section*{Acknowledgement}
The experiment was performed in the framework of the Super-FRS Collaboration for FAIR.
This work is partly supported by a Grant-in-Aid for Scientific Research on Innovative Areas (No. 24105705)
from the Ministry of Education, Culture, Sports, Science and Technology (MEXT), Japan,
and a Grant-in-Aid for Young Scientists (A) (No. 25707018)
from Japan Society for the Promotion of Science (JSPS).
%
%

\end{document}